\documentstyle[12pt]{article}
\input psfig.sty
\def\nn{\noindent}
\def\Re{{\cal R \mskip-4mu \lower.1ex \hbox{\it e}\,}}
\def\Im{{\cal I \mskip-5mu \lower.1ex \hbox{\it m}\,}}
\def\ie{{\it i.e.}}
\def\eg{{\it e.g.}}
\def\etc{{\it etc}}
\def\etal{{\it et al.}}

\def\sub#1{_{\lower.25ex\hbox{$\scriptstyle#1$}}}
\def\tev{\,{\ifmmode\mathrm {TeV}\else TeV\fi}}
\def\gev{\,{\ifmmode\mathrm {GeV}\else GeV\fi}}
\def\mev{\,{\ifmmode\mathrm {MeV}\else MeV\fi}}
\def\to{\rightarrow}

\def\subw{_{\rm w}}
\def\mh{\ifmmode m\sbl H \else $m\sbl H$\fi}
\def\mch{\ifmmode m_{H^\pm} \else $m_{H^\pm}$\fi}
\def\mt{\ifmmode m_t\else $m_t$\fi}
\def\mc{\ifmmode m_c\else $m_c$\fi}
\def\mz{\ifmmode M_Z\else $M_Z$\fi}
\def\mw{\ifmmode M_W\else $M_W$\fi}
\def\mws{\ifmmode M_W^2 \else $M_W^2$\fi}
\def\mhs{\ifmmode m_H^2 \else $m_H^2$\fi}   
\def\mzs{\ifmmode M_Z^2 \else $M_Z^2$\fi}
\def\mts{\ifmmode m_t^2 \else $m_t^2$\fi}
\def\mcs{\ifmmode m_c^2 \else $m_c^2$\fi}
\def\mchs{\ifmmode m_{H^\pm}^2 \else $m_{H^\pm}^2$\fi}
\def\ztwo{\ifmmode Z_2\else $Z_2$\fi}
\def\zone{\ifmmode Z_1\else $Z_1$\fi}
\def\mtwo{\ifmmode M_2\else $M_2$\fi}
\def\mone{\ifmmode M_1\else $M_1$\fi}
\def\tb{\ifmmode \tan\beta \else $\tan\beta$\fi}
\def\xw{\ifmmode x\subw\else $x\subw$\fi}
\def\ch{\ifmmode H^\pm \else $H^\pm$\fi}
\def\lum{\ifmmode {\cal L}\else ${\cal L}$\fi}
\def\inpb{\,{\ifmmode {\mathrm {pb}}^{-1}\else ${\mathrm {pb}}^{-1}$\fi}}
\def\infb{\,{\ifmmode {\mathrm {fb}}^{-1}\else ${\mathrm {fb}}^{-1}$\fi}}
\def\epem{\ifmmode e^+e^-\else $e^+e^-$\fi}
\def\ppb{\ifmmode \bar pp\else $\bar pp$\fi}
\def\bsg{\ifmmode B\to X_s\gamma\else $B\to X_s\gamma$\fi}
\def\bsll{\ifmmode B\to X_s\ell^+\ell^-\else $B\to X_s\ell^+\ell^-$\fi}
\def\bstt{\ifmmode B\to X_s\tau^+\tau^-\else $B\to X_s\tau^+\tau^-$\fi}
\def\lamt{\ifmmode \tilde\lambda\else $\tilde\lambda$\fi}
\def\shat{\ifmmode \hat s\else $\hat s$\fi}
\def\that{\ifmmode \hat t\else $\hat t$\fi}
\def\uhat{\ifmmode \hat u\else $\hat u$\fi}

\newskip\zatskip \zatskip=0pt plus0pt minus0pt
\def\matth{\mathsurround=0pt}

\def\atversim#1#2{\lower0.7ex\vbox{\baselineskip\zatskip\lineskip\zatskip
  \lineskiplimit 0pt\ialign{$\matth#1\hfil##\hfil$\crcr#2\crcr\sim\crcr}}}

\renewcommand{\thefootnote}{\fnsymbol{footnote}}

\begin{document} \begin{titlepage} 
\rightline{\vbox{\halign{&#\hfil\cr
&SLAC-PUB-8036\cr
&December 1998\cr}}}
\begin{center}

{\Large\bf More and More Indirect Signals for Extra Dimensions at More and 
More Colliders}
\footnote{Work supported by the Department of 
Energy, Contract DE-AC03-76SF00515}
\medskip

\normalsize 
{\large Thomas G. Rizzo } \\
\vskip .3cm
Stanford Linear Accelerator Center \\
Stanford University \\
Stanford CA 94309, USA\\
\vskip .3cm

\end{center}

\begin{abstract} 
It has been recently suggested by Arkani-Hamed, Dimopoulos and Dvali that 
gravity may become strong at energies not far above the electroweak scale and 
thus remove the hierarchy problem. Such a scenario can be tested at both 
present and future accelerators since towers of Kaluza-Klein gravitons and 
associated scalar fields now play an important phenomenological role. In this 
paper we examine several processes for their sensitivity to a low scale for 
quantum gravity including deep inelastic $ep$ scattering at HERA, high 
precision low energy $\nu N$ scattering, Bhabha and Moller scattering at 
linear colliders and both fermion and gluon pair production at 
$\gamma \gamma$ colliders. 
\end{abstract} 




\renewcommand{\thefootnote}{\arabic{footnote}} \end{titlepage}


\section{Introduction}

Arkani-Hamed, Dimopoulos and Dvali(ADD){\cite {nima}} have recently 
proposed a radical solution to the hierarchy problem, \ie, the problem of 
why the weak scale is so far removed from the Planck scale, $M_{pl}$, 
where gravity becomes as strong as the other forces. ADD 
hypothesize the existence of $n$ additional large spatial dimensions in 
which gravity can live, called `the bulk' whereas all of the fields of the 
Standard Model are constrained to lie on `the wall', which is a 3-dimensional 
brane and corresponding to our conventional 4-dimensional world. It has 
recently been shown that a scenario of this type may emerge in 
string models where the effective Planck scale in the bulk is identified 
with the string scale{\cite {nima,tye}}. 
That the SM fields must remain on the wall without being excited into the bulk  
below some mass scale of order of a few TeV is argued based on the 
well-known behavior of 
QED down to rather short distances, the lack of observation of degenerate 
mirror copies of the SM fields and the experimental value of the width of the 
$Z$ boson{\cite {nima}}. Thus in the ADD scenario, gravity only 
appears to be weak in our ordinary 4-dimensional space-time since we have up 
to now merely 
observed it's action on the wall. In such a theory the hierarchy can be 
simply removed by postulating that the string or effective Planck scale in 
the bulk, $M_s$, is not far above the weak scale, \eg, a few TeV. Gauss' Law 
then provides a link between the values of $M_s$, $M_{pl}$, and the size of 
the compactified extra dimensions, $R$, 
\begin{equation}
M_{pl}^2 \sim R^nM_s^{n+2}\,,
\end{equation}
where the constant of proportionality depends not only on the value of $n$ 
but upon the geometry of the compactified dimensions. Interestingly, if $M_s$ 
is near a TeV then $R\sim 10^{30/n-19}$ meters; within Newtonian gravity 
and for fixed $n$, $R$ can be thought of as a critical point in the power-law 
behavior for the force of gravity. For two masses separated by a distance 
greater than $R$ one obtains the usual $1/r^2$ force law; however, for 
separations smaller than $R$ the power law changes to $1/r^{2+n}$. For $n=1$, 
$R\sim 10^{11}$ meters and is thus obviously excluded, but, for $n=2$ one 
obtains $R \sim 1$~mm, which is at the very edge of the range of sensitivity 
for existing experiments{\cite {test}}. For $2<n \leq 7$, where 7 is 
the maximum value of $n$ 
being suggested by M-theory, the value of $R$ is further reduced and thus we 
may conclude that the range $2\leq n \leq 7$ is of phenomenological interest. 
While we feel the ADD scenario is quite compelling, we note that several other 
sets of authors have considered alternate models based on the suggestion of 
a low Planck or string scale within other contexts{\cite {contexts}} through 
the use of extra compactified dimensions. Only the ADD scenario will concern 
us in what follows.

The phenomenology of the ADD model as far as the new gravitational interactions 
are concerned can be obtained by considering a linearized theory of gravity 
in the bulk, decomposing it into the more familiar 4-dimensional states and 
recalling the existence of Kaluza-Klein towers for each of the conventionally 
massless fields. The entire set of fields in the K-K tower couples in an 
identical fashion to those of the SM. By considering the forms of the $4+n$  
symmetric conserved stress-energy tensor for the various SM fields and by 
remembering that such fields live only on the wall, the 
relevant Feynman rules can be derived{\cite {pheno}}. An important result of 
these considerations is that only the massive spin-2 K-K towers (which couple 
to the 4-dimensional stress-energy tensor, $T^{\mu\nu}$) and spin-0 K-K 
towers (which couple proportional to the trace of $T^{\mu\nu}$) are of 
phenomenological relevance as all the spin-1 fields can be shown to decouple 
from the particles of the 
SM. If the processes under consideration are at tree-level and 
involve only massless fermions and gauge fields, as will be the case below, 
the contributions of the spin-0 fields can also be safely ignored. There will, 
however, be other processes where these scalars play an important role.

Given the Feynman rules as developed in {\cite {pheno}} it appears that the 
ADD scenario has two basic classes of collider tests: ($i$) The emission of 
a (kinematically cut off) tower of gravitons during a hard collision 
leads to missing energy final states at either lepton or 
hadron colliders since the emitted gravitons essentially 
do not interact with the detector. The rate for such processes is quite 
sensitive to the value of $n$, falling rapidly as the number of 
dimensions increases beyond $n=2$. 
The advantage to such processes is that their observation together with a fit 
to the missing energy spectrum would tell us the value of $n$. The clear 
disadvantage is due to the rapid fall off in rate with large 
$n$ which makes the process difficult to observe above SM backgrounds in 
that case. ($ii$) The exchange of a K-K graviton 
tower between SM fields can lead to almost $n$-independent modifications to 
conventional cross sections and distributions or can possibly lead to new 
interactions such as $gg\to e^+e^-$ as discussed by Hewett{\cite {pheno}}. In a 
simple approximation the exchange of the graviton K-K tower leads to an 
effective operator of dimension-eight. Here 
one does not produce the gravitons directly and one does not learn much 
about the value of $n$ itself provided 
deviations attributable to gravity are indeed obtained experimentally. But this 
$n$-independence is also a strength since there is in this case 
no fall off in the size of the deviations with 
large $n$. For low $n$, both type-$i$ and type-$ii$ processes give comparable 
reach in sensitivity to the scale $M_s$ but, due to their approximate 
$n$-independence, type-$ii$ processes eventually 
win out{\cite {pheno}} for $n>2$.

In this paper we will extend the analyses of the ADD scenario as 
presented in {\cite {pheno}} to a set of previously unconsidered reactions of 
type-$ii$ in order to examine their sensitivity to values of 
$M_s$ of order a few TeV 
or less. In section 2, we extend the previous LEP/NLC and Tevatron/LHC 
studies{\cite {pheno}} to the case of 
neutral current interactions at HERA where K-K towers of gravitons are now 
exchanged in the $t$-channel during the $eq\to eq$ scattering process; 
it is important to note that such exchanges do not occur in the charged 
current channel since gravitons are both neutral as well as isoscalar. 
As is well known, the sensitivity of HERA to conventional 
dimension-six $eeqq$ contact interactions is both complementary and 
numerically comparable{\cite {contact}} to that obtainable from LEP and the 
Tevatron and hence a comparison of their $M_s$ sensitivity in the present case 
is particularly interesting. 
Such discussions naturally lead one to think about 
the potential sensitivity of high precision low energy 
$\nu N$ neutral current scattering experiments,  
such as NuTeV, who have recently{\cite {nu}} obtained a very competitive 
measurement of the $W$ mass (or the weak mixing angle) by employing the 
Paschos-Wolfenstein relation{\cite {pw}}. In Section 3 we will examine the 
sensitivity of these precise but relatively low-energy experiments to 
interesting values of $M_s$; unfortunately we find that while such processes 
are quite sensitive to dimension-six 
compositeness operators{\cite {contact,nu}}, 
there is little sensitivity to the string scale in this case. 
In section 4, we will return to a discussion of the 
$M_s$ sensitivity of various processes at lepton linear colliders by examining 
both Bhabha and Moller scattering, $e^\pm e^-\to e^\pm e^-$. It is often 
claimed that Moller 
scattering is the most sensitive of the purely leptonic processes accessible 
at lepton colliders to the existence of compositeness{\cite {tim}} and new 
neutral gauge bosons{\cite {cuypers}}. Thus it would appear natural to compare 
the sensitivity of these two processes to that obtained earlier by 
Hewett{\cite {pheno}} who examined the reactions $e^+e^-\to f\bar f$, 
$f\neq e$. These 
claims will be shown to indeed be valid for the case at hand when statistical 
errors are dominant. In Section 5 we will consider the $M_s$ sensitivity of 
the process 
$\gamma \gamma \to f\bar f$ via high energy $\gamma \gamma$ collisions 
obtainable at linear colliders through the backscattering of pairs of laser 
beams{\cite {telnov}}. Although the $M_s$ reach is somewhat lower here than 
in purely leptonic reactions, $\gamma \gamma \to f\bar f$ can provide 
complementary information. A summary of our analysis and our conclusions 
can be found in Section 6.

\section{HERA}

HERA is currently colliding 27.5 GeV electrons on 920 GeV protons, thus 
obtaining a center of mass energy of $\sqrt s=318$~GeV. Both the H1 and 
ZEUS experiments are expected{\cite {gayler} to collect $\sim 1~fb^{-1}$ in 
integrated luminosity over the next several years. After the year 2000, it is 
anticipated that HERA will deliver $\sim 60\%$ longitudinally polarized 
$e^\pm$ beams shared more or less equally between the four charge and 
polarization assignments. These specific lumonosity and polarization 
parameters will be assumed in our 
analysis below. We recall from the discussion above that we need only 
to consider neutral current processes since graviton towers are not 
exchanged at tree level in charged current reactions. Thus potential 
deviations in cross sections at 
high $Q^2$ appearing in {\it both} channels due to, \eg, leptoquarks, new 
gauge bosons or contact interactions {\it cannot} be attributed to the ADD 
model of low-scale quantum gravity. 

The basic subprocess cross section for $e^-_{L,R}q$ elastic 
scattering, now including the exchange of a K-K tower of gravitons, is given 
by {\cite {brw}} 
\begin{eqnarray}
{d\sigma_q \over dxdQ^2} & = &  {2\pi \alpha^2 \over {\hat s}^2}
\bigg[ {\rm SM}-C\bigg\{ \left( {Q_eQ_q\over t}
+{\sigma C'(v_e+\sigma a_e)v_q \over t-m_Z^2} \right) 2(u-\hat s)^3
\nonumber \\
&  & \quad -
{C'(a_e+\sigma v_e)a_q\over t-m_Z^2}\,t\,[t^2-3(u-t)^2]\bigg\}\nonumber \\
& &\quad +{C^2\over 2}\{t^4-3t^2(u-\hat s)^2+4(u-\hat s)^4\}\biggr]\,,
\end{eqnarray}
where `SM' is the conventional SM contribution, 
$C=\lambda K/(4\pi \alpha M_s^4)$, $C'=\sqrt 2 G_FM_Z^2/4\pi \alpha$ and 
$\sigma=\pm 1$ for left-(right-)handed electrons. 
We note here that through the use of crossing symmetry, this cross 
section with suitable modifications can be shown to reproduce those 
obtained for by Hewett and by Guidice, Rattazzi and Wells{\cite {pheno}} 
with the following caveat regarding the parameter $K$ in the expressions 
above. For $K=1(\pi/2)$ we 
recover the normalization convention employed by Hewett(Guidice, Rattazzi and 
Wells){\cite {pheno}}; we will take $K=1$ in the numerical analysis that 
follows but keep the factor in our analytical expressions. 
We recall from the Hewett analysis that $\lambda$ is 
a parameter of order unity whose sign is undetermined and that, given the 
scaling relationship between $\lambda$ and $M_s$, experiments in the 
case of processes 
of type-$ii$ actually probe only the combination $M_s/|K\lambda|^{1/4}$. 
For simplicity in what follows we will numerically set $|\lambda|=1$ and 
employ $K=1$ but we caution the reader about this 
technicality and quote our sensitivity to $M_s$ for $\lambda=\pm 1$. 

In the case of $e^-_{L,R}\bar q$ scattering, we simply let $a_q\to -a_q$ in the 
above expression and make the replacement $q(x)\to \bar q(x)$ in the 
sum over initial state partons. Here, and in the expression above 
$Q^2=-t=y\hat s=sxy$ and $u=-\hat s-t=-sx(1-y)$ with 
$Q^2,x,y$ being the conventional variables of deep inelastic scattering. 
For positron scattering we note the relations 
$d\sigma^+_{R,L}(q,\bar q)=d\sigma^-_{L,R}(\bar q,q)$ can be used to obtain 
the complementary cross sections. We note further that with the  
normalization employed above $a_e=-1/2$.

Of course in the ADD scenario, the $eq\to eq$ process is not the only one 
which contributes to deep inelastic scattering. Since both electrons and 
gluons have non-zero stress-energy tensors, a tower of K-K gravitons 
can also be exchanged in the $t$-channel mediating the process $eg\to eg$ 
where the squared matrix element 
is independent of the charge and helicity of the incoming lepton. 
The corresponding subprocess cross section for $e^\pm_{L,R}g$ scattering 
is thus relatively simple and is given by
\begin{equation}
{d\sigma_g \over {dxdQ^2}}={-\lambda^2 K^2\over {\pi M_s^8\hat s^2}}u\hat s
[(u^2+\hat s^2)]\,,
\end{equation}
there being no SM contribution in this case. Note that with $K$ taking on the 
values discussed above, using crossing symmetry and rearranging color factors, 
we reproduce the structure of the analogous cross section expressions given 
by Hewett and by Guidice, Rattazzi and Wells{\cite {pheno}}.

In order to gauge the HERA sensitivity to exchanges of a K-K tower of 
gravitons, we follow the current HERA analysis technique as presented by 
Stanco{\cite {contact}}. Since this new exchange only reveals itself at higher 
values of $Q^2$, we divide the $Q^2$ range into two regions: below 
$Q^2=1000$ GeV$^2$ we assume that the SM holds and use this regime to normalize 
the neutral current cross sections for the four charge/polarization states of 
the incoming lepton. This assumption will be explicitly validated in the 
discussion below. 
Above $Q^2=1000$ GeV$^2$ we divide the range into 17 
$Q^2$ bins up to the kinematic limit; the location and width of these bins are 
essentially those of the present HERA analyses with only minor modifications 
due to the higher anticipated integrated luminosities. We then use a toy 
Monte Carlo approach to generate `data' assuming a given integrated luminosity 
for each of the four charge/polarization states. These data are then fit to 
the $M_s$-dependent cross section to obtain a lower bound on $M_s$ at the 
$95\%$ CL. In performing this analysis we employ the CTEQ4M parton density 
distributions{\cite {cteq}} although our results are not sensitive to this 
particular choice. We assume that the potential of any large systematic error 
associated with the calorimeter energy scale can be avoided in obtaining 
these results. 

\vspace*{-0.5cm}
\nn
\begin{figure}[htbp]
\centerline{
\psfig{figure=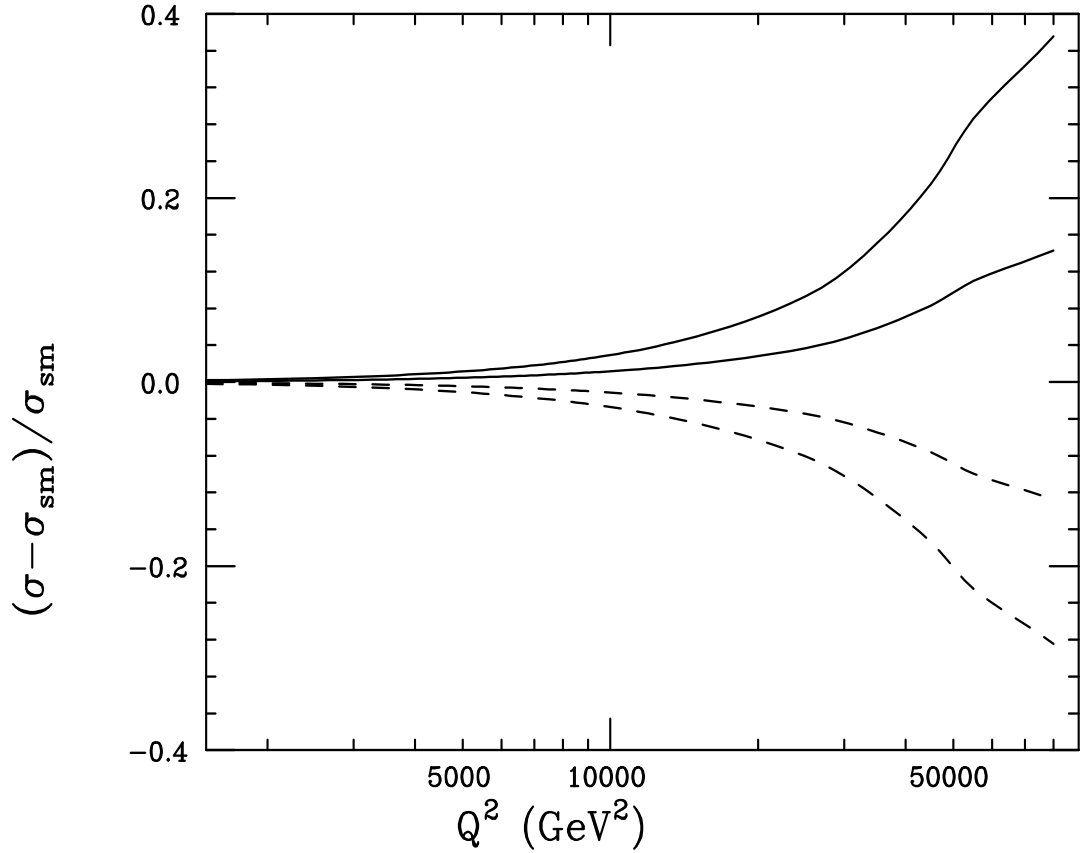,height=14cm,width=17cm,angle=-90}}
\vspace*{-1cm}
\caption[*]{Deviation in equally weighted sum of the $\sigma(e_{L,R}^\pm p)$ 
deep inelastic cross sections as a function of $Q^2$ for $\lambda=$1(solid) 
and -1(dashed). The outer(inner) curve in each case corresponds to assuming 
$M_s=800(1000)$ GeV.}
\label{fig1}
\end{figure}
\vspace*{0.4mm}

In examining the sensitivity of the four cross sections, 
$d \sigma(e_{L,R}^\pm p)$, one finds that the process with the 
largest(smallest) cross section (hence the best statistics) is the one with the 
least(most) sensitivity to $M_s$. Instead of trying to choose the beam that 
maximizes sensitivity to $M_s$ with the best statistics we will simply assume 
equal integrated luminosities are supplied for all four cases and combine the 
result into a single fit. One may either try to simultaneously 
fit to all four $e_{L,R}^\pm$ 
cross sections, \ie, $4\times 17$ bins, or simply fit to the sum of 
the four cross sections 
together in each $Q^2$ bin, \ie, 17 bins only. Given the need to have as much 
statistics as possible in the highest $Q^2$ bins we follow the latter approach. 
To get an idea of the resulting sensitivity we show in Fig.1 the deviation 
from the bin-integrated SM cross section for $M_s=$800 and 1000 GeV 
with $\lambda=\pm 1$. Note that the deviations from the SM grows only very 
slowly with increasing $Q^2$ and are not significantly noticeable below 
$Q^2=10000-15000$ GeV$^2$. 

\vspace*{-0.5cm}
\nn
\begin{figure}[htbp]
\centerline{
\psfig{figure=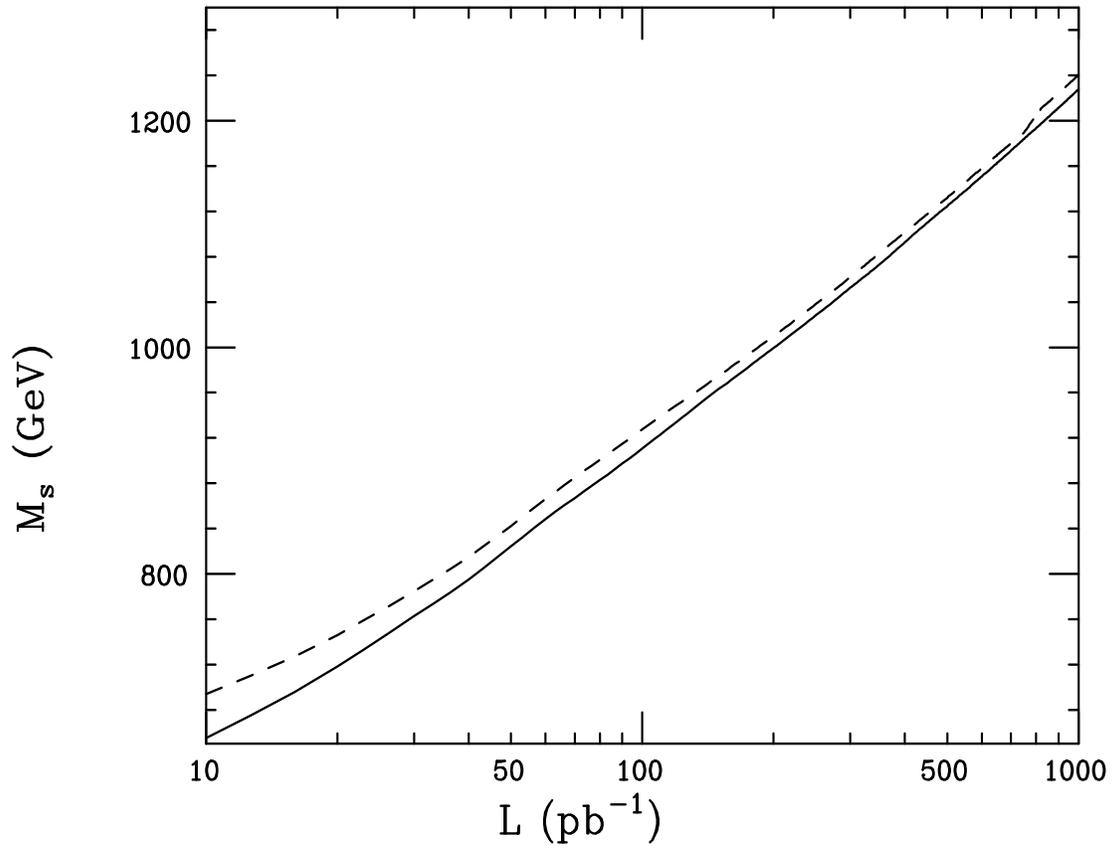,height=14cm,width=17cm,angle=-90}}
\vspace*{-1cm}
\caption[*]{$95\%$ CL lower bound on the value of $M_s$ obtainable at HERA as 
a function of the integrated luminosity per charge/polarization state for 
$\lambda=\pm 1$.}
\label{fig2}
\end{figure}
\vspace*{0.4mm}

Performing the analysis described above we arrive at the $95\%$ CL lower bound 
on $M_s$ as a function of the integrated luminosity as shown in Fig.2. With 
the assumption that each of the four neutral current processes, 
$\sigma(e_{L,R}^\pm p)$, obtain the same beam flux, the full 1 $fb^{-1}$ 
HERA luminosity corresponds to $L=250~pb^{-1}$ in this figure. The limit in 
this case, for either sign of $\lambda$, is $\simeq 1.04$ TeV which is very 
comparable to the potential search reach of 1.14 TeV obtainable at 
LEP II from the analysis of Hewett{\cite {pheno}}. Similarly it it comparable 
to, but somewhat lower than, that obtainable at Run II of the Tevatron 
through an analysis of the the Drell-Yan process. Clearly the bounds 
obtainable at HERA are complementary to those obtainable at other currently 
existing colliders. 

What limits are obtainable from the existing HERA data, \ie, approximately 
40 $pb^{-1}$ of integrated luminosity per experiment using an unpolarized 
$e^+$ beam at a center of mass energy of $\sqrt s \simeq 298$ GeV? The 
reduced center of mass energy and the use of an unpolarized $e^+$ beam both 
act to significantly suppress the reach relative to the estimate one would 
obtain by the use of the results shown in Fig.2 alone. We estimate that the 
current lower bound on $M_s$ from HERA to be no larger than $\simeq 500-600$ 
GeV.

\section{Low Energy $\nu$ Scattering}

Do lower energy measurements reveal anything about $M_s$? Since the exchange 
of K-K towers of gravitons is essentially flavor independent and is 
a parity conserving process these new effects will not show 
themselves in atomic parity violation or polarized lepton nucleon 
scattering experiments. The only other possibility is neutrino-nucleon 
neutral current deep inelastic scattering.

In the case of $\nu(\bar \nu)q$~and $\nu(\bar \nu)\bar q$  scattering we can 
obtain the relevant cross sections from the expressions above by setting 
$Q_e=0$, $v_e=a_e=1/2$, taking $Q^2 \ll M_Z^2$, and recalling that 
$\nu(\bar \nu)$'s are always left-(right-)handed. We then arrive at the 
following expression for the $\nu(\bar \nu)q$ subprocess cross section:
\begin{eqnarray}
{d\sigma_q^{\nu,\bar \nu}\over dxdy}&=&{G_F^2s\over 4\pi}\bigg[\{(v_q\pm a_q)^2+
(v_q\mp a_q)^2(1-y)^2\}+xF\{-2(2-y)^3v_q\nonumber \\
&\pm& y(y^2-3(2-y)^2)a_q\}+{(xF)^2\over 2}\{y^4-3y^2(2-y)^2+
4(2-y)^4\}\bigg]\,,
\end{eqnarray}
where $F=\lambda Ks/{\sqrt 2} G_FM_s^4$ and the upper(lower) sign is for the 
$\nu(\bar \nu)$ scattering process. The first term is just that 
arising from the SM while the additional terms arise from the K-K graviton 
tower exchange and its interference with the SM $Z$ exchange. The corresponding 
$\bar q$ cross section can be obtained by letting $a_q\to -a_q$ in the above 
expression. The corresponding 
$\nu(\bar \nu)g$ subprocess cross section which has a pure graviton exchange 
and no SM contribution is identical in both cases and is given by
\begin{equation}
{d\sigma_g^{\nu,\bar \nu}\over dxdy}={G_F^2s\over 4\pi}~8(xF)^2(1-y)
\{1+(1-y)^2\}\,.
\end{equation}
To obtain the complete scattering cross section one must weight the two 
expressions above with the relevant parton density functions(PDFs):
\begin{equation}
{d\sigma^{\nu,\bar \nu}\over dxdy}= \sum_q 
\{{d\sigma_q^{\nu,\bar \nu}\over dxdy} xq(x)
+{d\sigma_{\bar q}^{\nu,\bar \nu}\over dxdy} x\bar q(x)\}
+{d\sigma_g^{\nu,\bar \nu}\over dxdy} xg(x)\,,
\end{equation}
with the sum extending over all quark flavors. 

To get an idea of the 
sensitivity of neutrino nucleon scattering to the exchange of K-K towers of 
gravitons it is instructive to form the well known ratios
$R^{\nu,\bar \nu}=\sigma_{NC}^{\nu,\bar \nu}/\sigma_{CC}^{\nu,\bar \nu}$ 
for an isoscalar target in the valence quark approximation which then allows 
us to trivially perform the integrations over the $x$ and $y$ variables. (We 
note that these quantities are not quite as well measured{\cite {pdg}} as is 
the Paschos-Wolfenstein relation to be discussed later below). 
We obtain the expressions 
\begin{eqnarray}
R^\nu &=& g_L^2(u)+g_L^2(d)+{1\over 3}g_R^2(u)+{1\over 3}g_R^2(d)
+\Delta,\nonumber \\
R^{\bar \nu} &=& g_L^2(u)+g_L^2(d)+3g_R^2(u)+3g_R^2(d)+3\Delta\,,
\end{eqnarray}
where $g_L(u)=1/2-4/3\sin^2 \theta_w$, \etc., ~and $\Delta$ can be expressed 
numerically as
\begin{equation}
\Delta=-3.39\times 10^4F'R_1+2.15\times 10^{10}F'^2R_2+9.80\times 10^9
F'^2R_3\,,
\end{equation}
where $F'=\lambda Ks/M_s^4$ (with $\sqrt s$ and $M_s$ in GeV) and the 
$R_i$ are ratios of integrals over the appropriate PDFs:
\begin{eqnarray}
R_{1,2}&=&{{\mbox {$\int x^{2,3}[u(x)+d(x)]~dx$}}\over {\mbox {$\int x[u(x)
+d(x)]~dx$}}}\nonumber \\
R_{3}&=&{{\mbox {$\int 2x^3g(x)~dx$}}\over {\mbox {$\int x[u(x)+d(x)]~dx$}}}\,,
\end{eqnarray}
which need to be evaluated at a typical value of 
$Q^2$ and over the relevant $x$ range 
for a given experiment. For a typical $Q^2$ of 25 GeV$^2$ and $0.001<x<1$ we 
find, using the CTEQ4M PDFs{\cite {cteq}}, that $R_1\simeq 0.21$, 
$R_2\simeq 0.071$, and $R_3\simeq 0.042$. Since the $R_i$ are not too small 
and the numerical coefficients in Eq.(8) are large, one might anticipate a 
reasonable sensitivity to the string scale $M_s$. However, a short analysis 
shows this not to be the case due to the low values of the center of mass 
energy obtained in such collisions. Although the peak neutrino energies at 
NuTeV may be as high at 400 GeV, the average energies of the $\nu_\mu$ and 
$\bar \nu_\mu$ from the Fermilab Tevatron Quadrupole triplet neutrino beam are 
roughly 165 and 135 GeV, respectively{\cite {nu}}, implying that the typical 
$\sqrt s$ for these collisions is only $\simeq 17$ GeV. In turn, assuming 
$K=1$, we arrive at $F' \simeq 3\times 10^{-10}$ and thus 
$\Delta \simeq -2.15\times 10^{-6}$ which is far too small to be observable 
at any forseeable level of precision. Note that this value 
would only be an order of 
magnitude larger if all neutrinos in the beam had their maximum possible 
energies. 

Next, from the considerations above we are able to directly construct the 
Paschos-Wolfenstein relationship for an isoscalar target. We anticipate that 
this now will take the more general form including the effects 
of sea quarks (since they cancel in the differences in both the numerator and 
denominator) but neglecting charm mass effects, 
\begin{equation}
R_{PW}={\sigma_{NC}^\nu -\sigma_{NC}^{\bar \nu}\over \sigma_{CC}^\nu-\sigma_
{CC}^{\bar \nu}}={1\over 2}-\sin^2 \theta_w+\Delta'\,,
\end{equation}
where $\Delta'$ arises from graviton exchange and its interference with the SM 
amplitude. 
Note these K-K contributions only appear in the numerator of the 
above expression. Several things are immediately obvious. First the 
$\nu g\to \nu g$ and $\bar \nu g\to \bar \nu g$ contributions, being the same, 
cancel as do those corresponding to the pure graviton terms in the difference 
between $\nu q(\bar q)\to \nu q(\bar q)$ and $\bar \nu q(\bar q)\to 
\bar \nu q(\bar q)$. Secondly, the term proportional to the parity conserving 
vector coupling of 
the quarks, $v_q$, in the SM-graviton interference term will also cancel 
with the only remaining term being proportional to $a_q$. This leaves us, after 
integration over $y$, with the result
\begin{equation}
R_{PW}=\sum_V v_qa_q -{15\over 8}F{\int \sum_V a_q x^2 q_V(x)~dx\over 
\int \sum_V x q_V(x)~dx}\,,
\end{equation}
where the sum extends over the valence partons in the isoscalar target. The 
first term once expanded in terms of the conventional $Z$ boson couplings is 
just that provided by the SM while the second SM-graviton interference term, 
$\Delta'$, can be 
shown to vanish! Since $u_V(x)=d_V(x)$ in an isoscalar target and $a_u=-a_d$ 
the sum in the numerator is identically zero. This result tells us that 
K-K gravitons do not influence the Paschos-Wolfenstein relation whatsoever, 
something we may have expected due to their isoscalar nature. 

It appears that low energy neutrino measurements, however precise, will not 
tell us much, if anything, about the scale $M_s$. 
One may ask why $\nu N$ scattering is sensitive to traditional contact 
interactions but not to the exchange of a K-K tower of gravitons. The answer 
is directly related to the fact that traditional contact interactions are 
dimension-six operators while those induced by low scale quantum gravity are 
dimension-eight. With coefficients of order unity, a scale of order 1 TeV 
and an average $\sqrt s=$17 GeV, the dimension-eight operators are suppressed 
relative to those of dimension-six by a factor of $\simeq 3500$! For these 
dimension-eight operators the high precision of the data cannot offset their 
being at rather low energies. To search for 
$M_s$ in the ADD scenario we clearly need larger collision energies than 
those provided by $\nu N$ scattering.

\section{Bhabha and Moller Scattering at Linear Colliders}

Linear colliders will provide the opportunity to make precision measurements 
of a number of elementary processes in the $\sqrt s=500-1500$ GeV energy 
range. In addition to the conventional processes $e^+e^-\to f\bar f$, whose 
sensitivity to the exchange of a K-K tower of gravitons was discussed by 
Hewett{\cite {pheno}}, both Bhabha and Moller scattering offer complementary 
opportunities. In principle, Moller scattering, which takes place at a 
future linear collider run in the $e^-e^-$ mode{\cite {emem}}, may be of 
particular interest due to its well-known sensitivity to both contact 
interactions and $Z'$ exchange{\cite {tim,cuypers}}.

In analyzing both the Bhabha and Moller processes we will make an angular 
acceptance cut of $10^o$ with respect to the incoming beams, assume a $90\%$ 
$e^-$ beam polarization $P$, with an uncertainty of 
$\delta P/P=0.3\%${\cite {zdr}} and a integrated luminosity uncertainty of 
$\delta L/L=0.1\%${\cite {bw}}. (We will ignore the possibility of polarizing 
the positron beam in the present analysis.) In the case of Moller scattering 
both $e^-$ beams are assumed to have identical polarization so that the 
{\it effective} beam polarization will be $P_{eff}=2P/(1+P^2)\simeq 0.9945$ 
with a 
correspondingly decreased uncertainty of $\delta P_{eff}/P_{eff}
\simeq 0.032\%$. In the subsequent analysis the effects of initial 
state radiation will be included in all processes 
and we will assume a lepton identification efficiency of  $100\%$.

\vspace*{-0.5cm}
\nn
\begin{figure}[htbp]
\centerline{
\psfig{figure=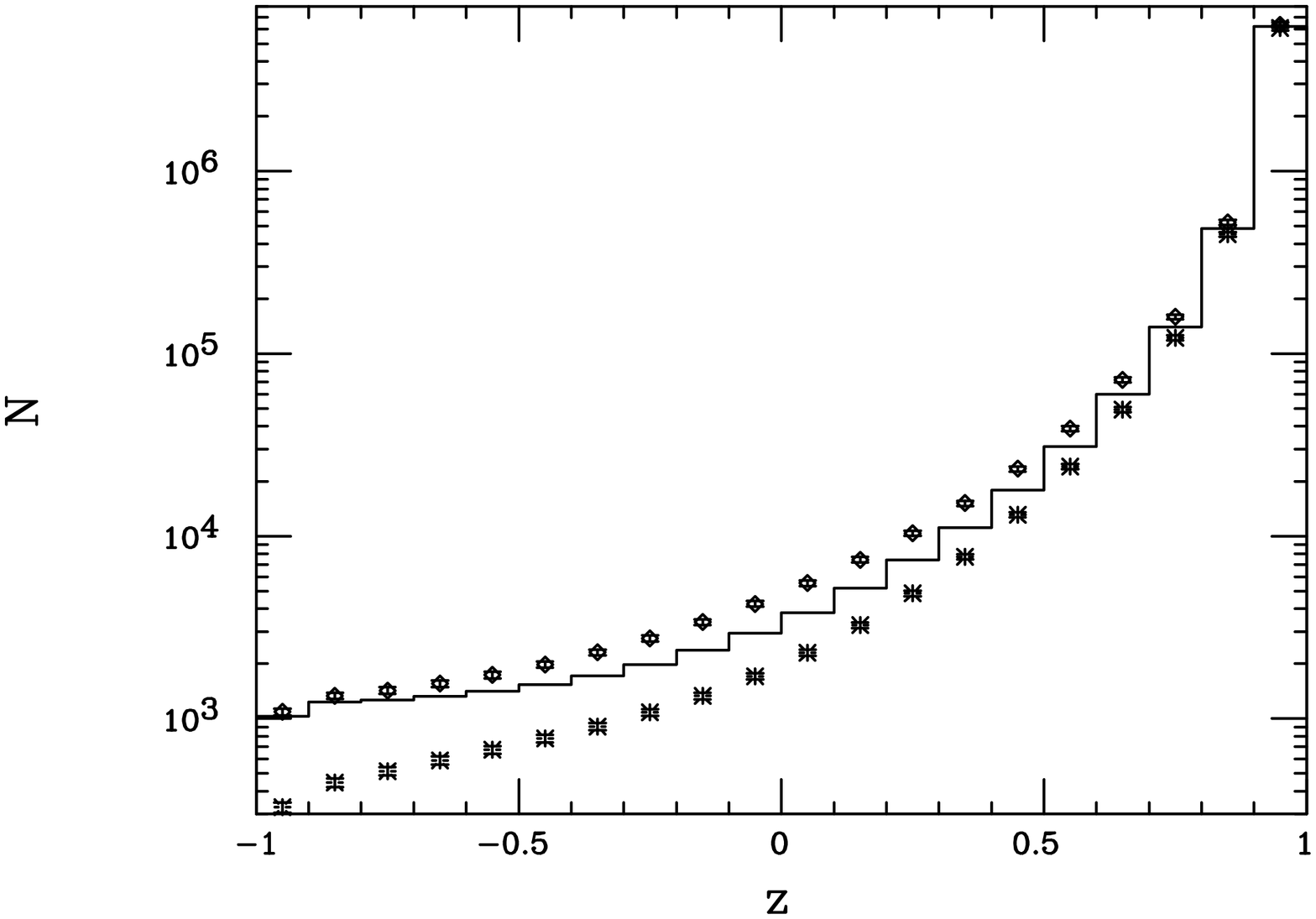,height=10.5cm,width=14cm,angle=-90}}
\vspace*{-10mm}
\centerline{
\psfig{figure=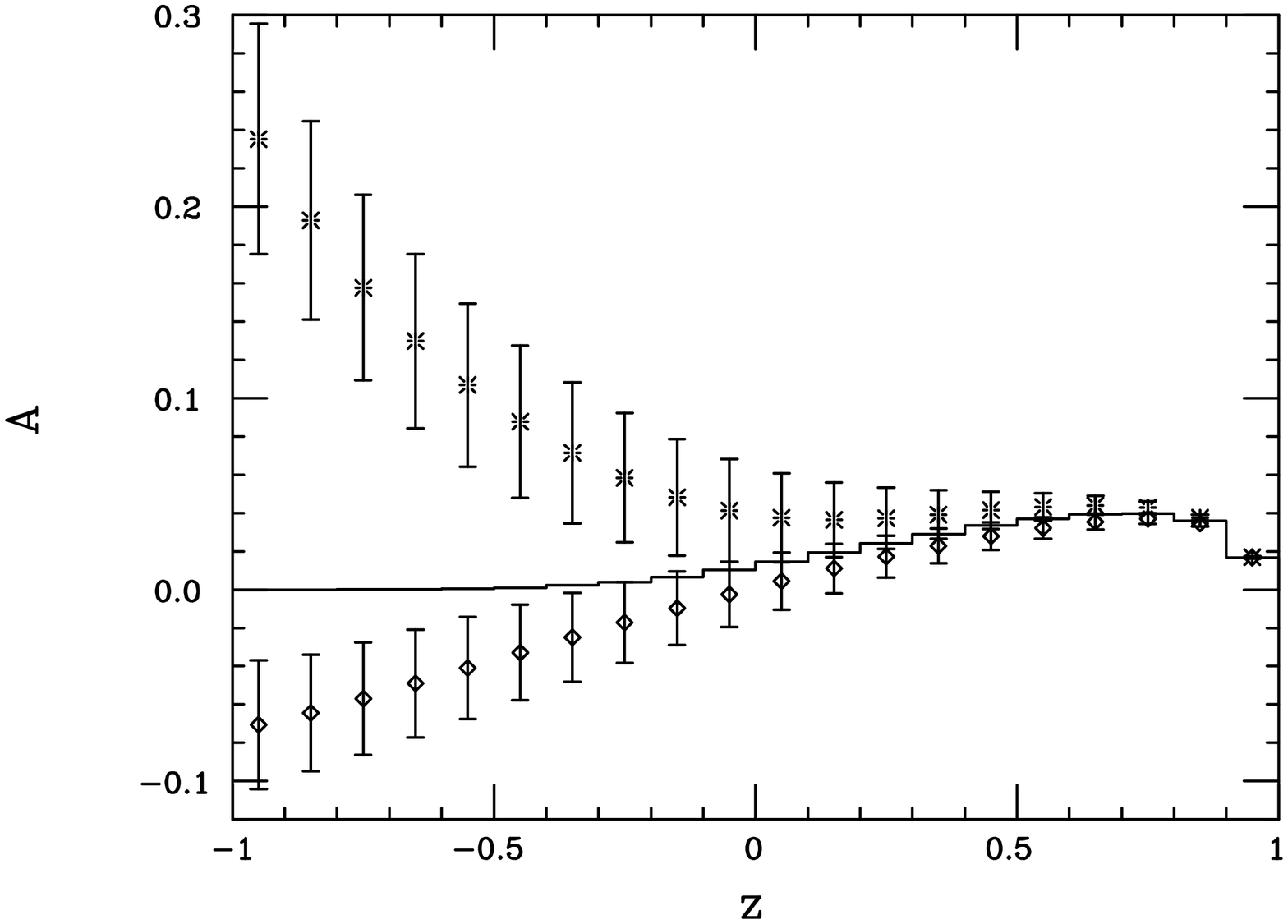,height=10.5cm,width=14cm,angle=-90}}
\vspace*{-0.9cm}
\caption{Deviation from the expectations of the SM(histogram) for Bhabha 
scattering at a 500 GeV 
$e^+e^-$ collider for both the(top) number of events per angular bin, $N$, 
and the Left-Right polarization asymmetry(bottom) as a function of 
$z=\cos \theta$ assuming $M_s$=1.5 TeV. The two sets of data points 
correspond to the choices $\lambda=\pm$ 1 and an assumed integrated 
luminosity of $75~fb^{-1}$.}
\label{fig3}
\end{figure}
\vspace*{0.4mm}

In the case of Bhabha scattering the differential cross section can be 
written as
\begin{eqnarray}
{d\sigma_B \over {dz}} &=& {\pi \alpha^2 \over {s}}\bigg[{\rm SM}-2C
\bigg\{F_1(s,t) \nonumber \\
& &\quad + \bigg[{F_2(s,t)v_e^2+F_3(s,t)a_e^2\over {(s-M_Z^2)}}+
(s\leftrightarrow t)\bigg]\bigg\}+C^2 F_4(s,t)\bigg]\,,
\end{eqnarray}
where `SM' in the expression above now corresponds to the usual SM contribution 
to Bhabha scattering, $z=\cos \theta$, 
$C=\lambda K/(4\pi \alpha M_s^4)$ as in the expressions above 
and the kinematic functions $F_i$ are given by
\begin{eqnarray}
F_1(s,t)&=& 9({s^3\over {t}}+{t^3\over {s}})+23(s^2+t^2)+30st\,, \nonumber \\
F_2(s,t)&=& 5s^3+10s^2t+18st^2+9t^3\,, \nonumber \\
F_3(s,t)&=& 5s^3+15s^2t+12st^2+t^3\,, \nonumber \\
F_4(s,t)&=& 41(s^4+t^4)+124st(s^2+t^2)+148s^2t^2\,.
\end{eqnarray}
Employing finite beam polarization the corresponding angular-dependent 
polarized Left-Right Asymmetry can be expressed as 
\begin{equation}
A_{LR}={{\biggl[{\rm SM'}-2Cv_ea_e\bigg\{{F_2(s,t)+F_3(s,t)\over {(s-M_Z^2)
}}+(s\leftrightarrow t)\bigg\}\biggr]\over {\biggl[{\rm SM}-2C\bigg\{F_1(s,t)
+\bigg[{F_2(s,t)v_e^2+F_3(s,t)a_e^2\over {(s-M_Z^2)}}+(s\leftrightarrow t)
\bigg]\bigg\}+C^2 F_4(s,t)\biggr]}}}\,. 
\end{equation}

Given these expressions we can obtain the search 
reach for $M_s$ for a given integrated luminosity using the assumptions 
discussed above by fitting to the total number of events, the shape of the 
angular distribution and the angle-dependent values of $A_{LR}$. We divide 
the angular range into 20 equal-sized $\cos \theta$ bins of width 
$\Delta z=0.1$, except for those nearest the beam pipe due to the above 
mentioned cut. To first get an idea of the influence of finite $M_s$ 
we show the distributions for Bhabha scattering in Fig. 3 for the 
case of a $\sqrt s$=500 GeV lepton collider with an integrated luminosity of 
75 $fb^{-1}$ assuming $M_s=1.5$ TeV. In this figure the cross section in 
the forward direction is dominated by the photon pole but significant 
deviations from the SM, which is represented as 
the histogram, are observed away from this 
region in both the angular distribution and the Left-Right Asymmetry. Note 
the huge statistics available here. The 
two sets of data points show the size of the anticipated errors for both 
$\lambda=\pm 1$; note that they are mutually distinguishable. It is clear 
from this figure that for this center of mass energy and integrated 
luminosity the 
discovery reach for $M_s$ will be significantly larger than 1.5 TeV. 

\vspace*{-0.5cm}
\nn
\begin{figure}[htbp]
\centerline{
\psfig{figure=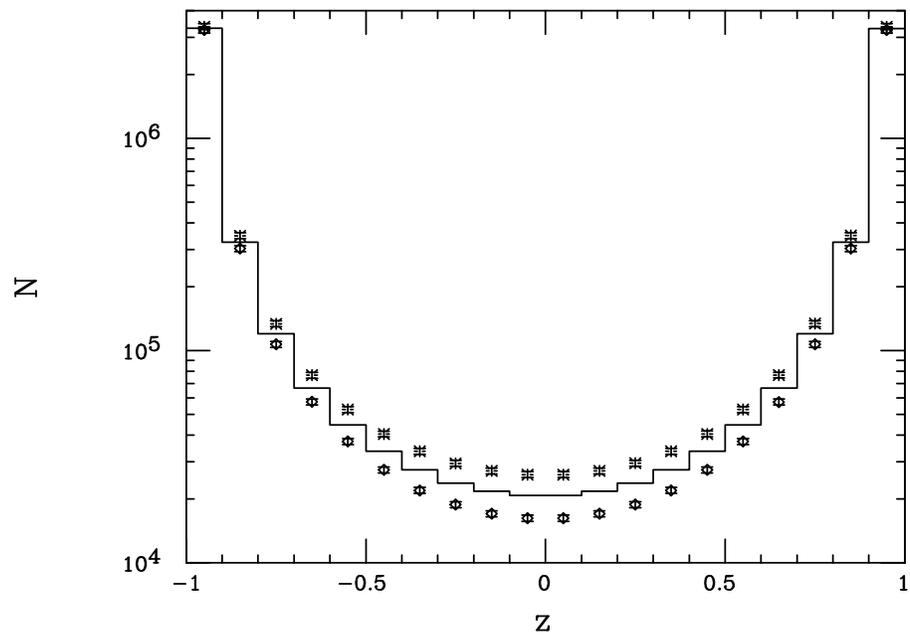,height=10.5cm,width=14cm,angle=-90}}
\vspace*{-10mm}
\centerline{
\psfig{figure=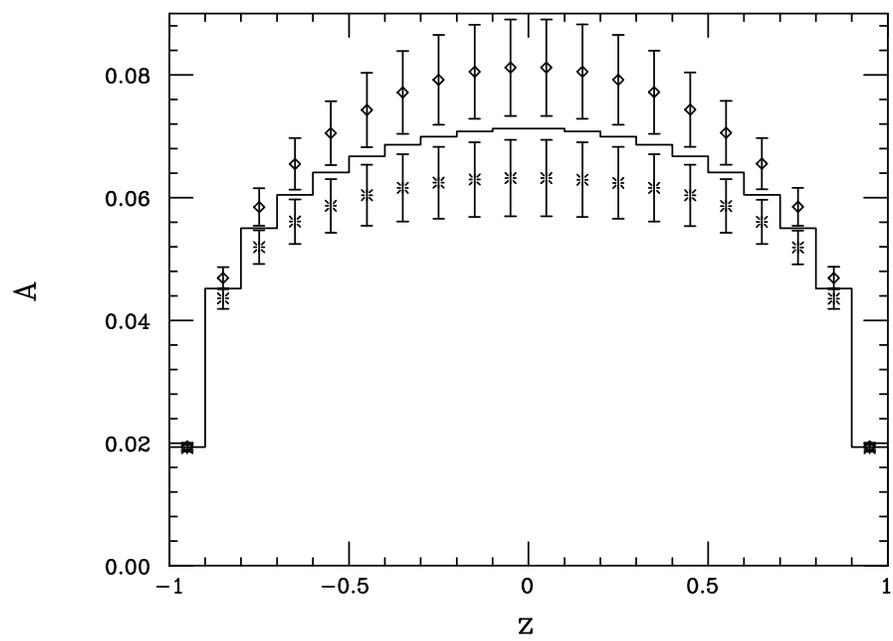,height=10.5cm,width=14cm,angle=-90}}
\vspace*{-0.9cm}
\caption{Same as the previous figure but now for Moller scattering.}
\label{fig4}
\end{figure}

In the case of Moller scattering one finds results similar to Bhabha 
scattering for both the cross section and Left-Right polarization asymmetry 
which can be obtained by 
crossing symmetry except for the overall factor of 2 in the normalization of
the cross section:
\begin{eqnarray}
{d\sigma_M \over {dz}} &=& {\pi \alpha^2 \over {2s}}\bigg[{\rm SM}-2C
\bigg\{F_1(u,t) \nonumber \\
& &\quad +\bigg[{F_2(u,t)v_e^2+F_3(u,t)a_e^2\over {(u-M_Z^2)}}+
(u\leftrightarrow t)\bigg]\bigg\}+C^2 F_4(u,t)\bigg]\,.
\end{eqnarray}
Note that the kinematic functions $F_i$ are now functions of $t$ and $u$ 
instead of $t$ and $s$ as in the case of Bhabha scattering. The corresponding 
expression for the polarized Left-Right Asymmetry is given by 
\begin{equation}
A_{LR}={{\biggl[{\rm SM'}-2Cv_ea_e\{{F_2(u,t)+F_3(u,t)\over {(u-M_Z^2)
}}+(u\leftrightarrow t)\}\biggr]\over {\biggl[{\rm SM}-2C\{F_1(u,t)
+\bigg[{F_2(u,t)v_e^2+F_3(u,t)a_e^2\over {(u-M_Z^2)}}+(u\leftrightarrow t)
\bigg]\}+C^2 F_4(u,t)\biggr]}}}\,.
\end{equation}

To get an idea of the sensitivity from Moller scattering we show in Fig.4 the 
results of the same analysis as presented in Fig.3. While the photon poles 
dominate both the forward and backward directions the central regions of both 
the angular distribution and the Left-Right Asymmetry show clear deviations 
from SM expectations. We again note the huge statistics that are available. 
However note that the overall deviation from the SM 
is perhaps not as great as in the case of Bhabha scattering due to there 
being 2 QED poles. Of course the extra pole also leads to increased 
statistics. Clearly the 
search reach for Moller scattering exceeds 1.5 TeV for this center of mass 
energy and integrated luminosity.

\vspace*{-0.5cm}
\nn
\begin{figure}[htbp]
\centerline{
\psfig{figure=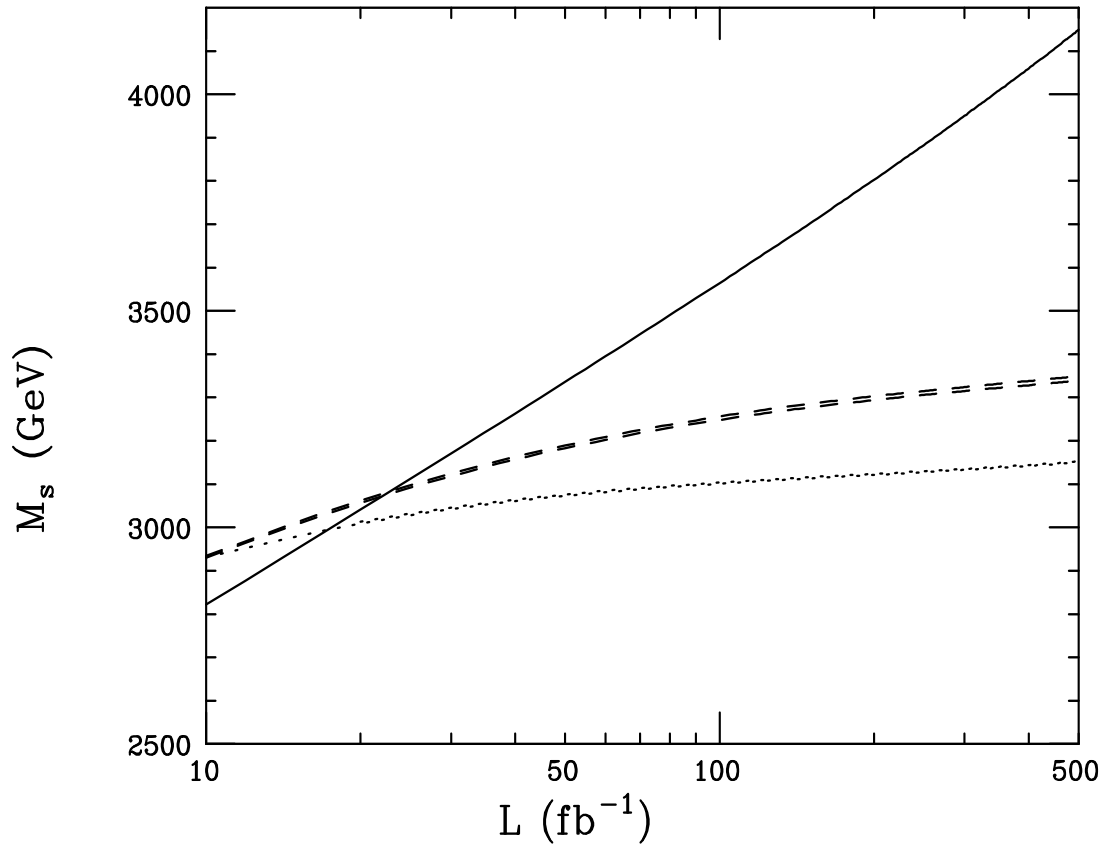,height=14cm,width=17cm,angle=-90}}
\vspace*{-1cm}
\caption[*]{Search reaches for $M_s$ at a 500 GeV $e^+e^-/e^-e^-$ collider as 
a function of the integrated luminosity for Bhabha(dashed) and Moller(dotted) 
scattering for either sign of the parameter $\lambda$ in comparison to the 
`usual' search employing $e^+e^-\to f\bar f$(solid) as described in the text.}
\label{fig5}
\end{figure}
\vspace*{0.4mm}

Figs. 5, 6 and 7 show the search reaches for 
$M_s$ as a function of the collider 
integrated luminosity for both Bhabha and Moller scattering in comparison 
to the `usual' search employing $e^+e^-\to f\bar f$ at $\sqrt s=500$ GeV, 
1 TeV and 1.5 TeV colliders, respectively. (In all three cases the results for 
$\lambda=\pm 1$ are shown but may not be visually separable.) We note that our 
result for the `usual' search confirms that of Hewett{\cite {pheno}} but 
is slightly higher due a different choice of angular cuts and assumed 
uncertainty of the integrated luminosity. Several results are immediately 
obvious from these two figures. First, for reasonable integrated luminosities, 
the search reaches for all three modes can exceed $\simeq 6\sqrt s$, which is 
rather remarkable. At a $\sqrt s=1.5$ TeV collider with a high integrated 
luminosity we see that string scales as high as 10 TeV can be probed. 
Second, since the traditional 
$e^+e^-\to f\bar f$ search with $f=\mu, ~\tau, ~b, ~c, ~t$, \etc. sums over 
many final states and employs many observables it tends to lead to the best 
search reach for most integrated luminosities, in particular, when large 
luminosity samples are available. In almost all cases the 
precision of this data is statistics dominated since there are only several 
thousands of events for each flavor. Third, the errors on the data in the cases 
of both Bhabha and Moller 
scattering are likely to be systematics dominated at typical integrated 
luminosities due to the huge event rates observed in Figs. 3 and 4. This 
explains the far shallower slopes of their luminosity dependence 
observed for both the Bhabha and Moller 
curves in these figures. Furthermore, for a fixed integrated luminosity, we 
know that three event rates for all three reactions decrease with increasing 
values of $\sqrt s$ leading to different weights in the errors between 
statistical and systematic. 
Thus we note, particularly in the case of Figs. 6 and 
7, that for low luminosities, where systematic errors are not as important as 
statistical ones, Moller scattering indeed leads to the best search reach for 
$M_s$ due to the huge statistics in that data sample in comparison to either 
Bhabha scattering or the conventional fermion pair 
channel. (Thus the explanation for why Bhabha scattering is a close 
second to Moller scattering in the search reach for $M_s$ for low luminosities 
becomes immediately obvious.) It is clear from this analysis that we again 
find complementarity in the search for TeV scale $M_s$ in the ADD scenario.

\vspace*{-0.5cm}
\nn
\begin{figure}[htbp]
\centerline{
\psfig{figure=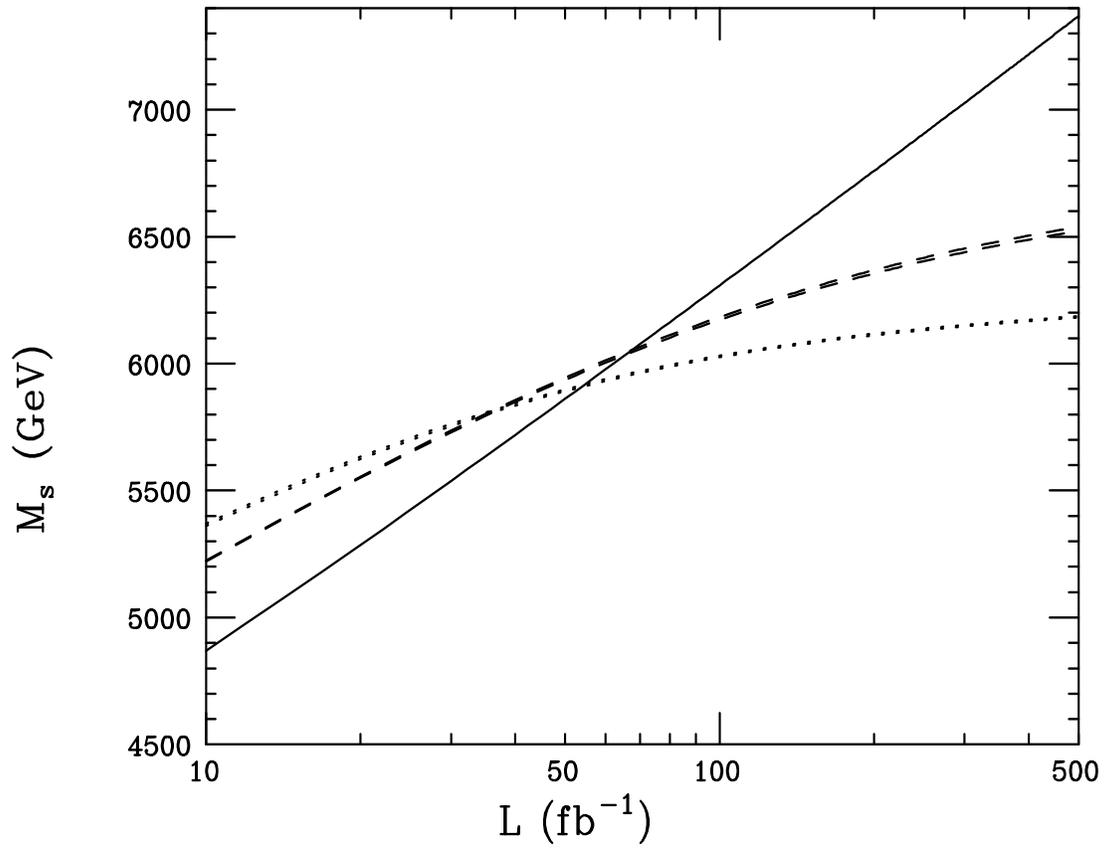,height=14cm,width=17cm,angle=-90}}
\vspace*{-1cm}
\caption[*]{Same as the previous figure but now for an $e^+e^-/e^-e^-$ 
collider with a center of mass energy of 1 TeV.}
\label{fig6}
\end{figure}
\vspace*{0.4mm}
\vspace*{-0.5cm}
\nn
\begin{figure}[htbp]
\centerline{
\psfig{figure=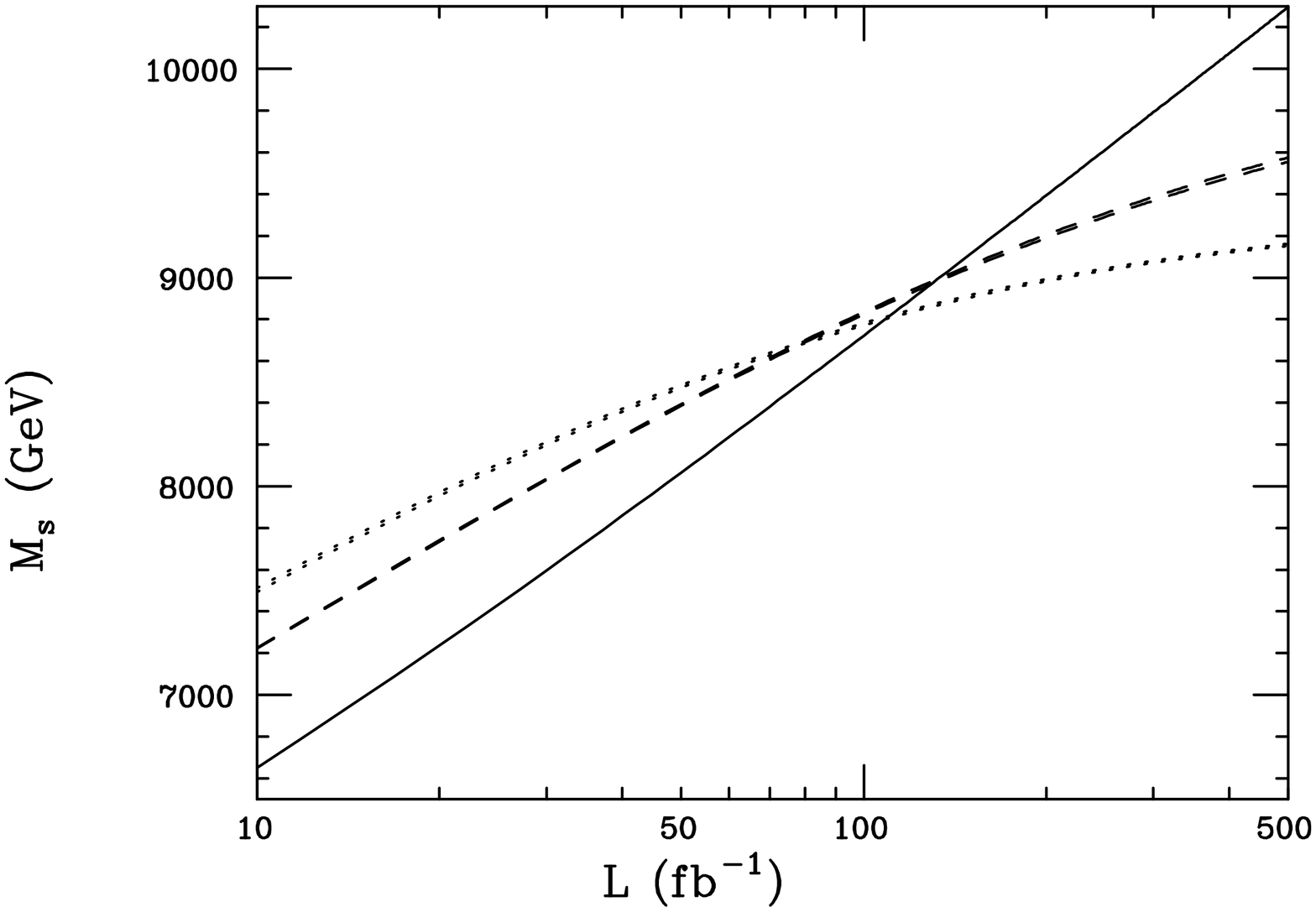,height=14cm,width=17cm,angle=-90}}
\vspace*{-1cm}
\caption[*]{Same as the previous figure but now for an $e^+e^-/e^-e^-$ 
collider with a center of mass energy of 1.5 TeV.}
\label{fig6n}
\end{figure}
\vspace*{0.4mm}

\section{$\gamma \gamma$ Colliders}

The process $\gamma \gamma \to f\bar f$ is 
particularly clean, there being no tree level corrections from electroweak 
effects, and has a long tradition as a probe for higher dimensional 
operators. In fact, no gauge invariant operators due to contact interaction 
exist at dimension-six. 

$\gamma \gamma$ collisions may be possible at future 
$e^+e^-$ linear colliders by the use of Compton backscattering of low energy 
laser beams{\cite {telnov}}. The backscattered laser photon spectrum, 
$f_\gamma(x={E_\gamma \over E_e})$, is far from being monoenergetic and is 
cut off above $x_{max}\simeq 0.83$ implying that the photons are significantly 
softer than their parent lepton beam energy. As we will see, this cutoff at 
large $x$, $x_{max}$, implies that the $\gamma \gamma$ center of mass energy 
never exceeds $\simeq 0.83$ of the parent collider and this will result in a 
significantly degraded $M_s$ search reach. We will ignore the possibility of 
employing polarized photon collisions in what follows but one would 
anticipate that the search reach would somewhat 
increase beyond what we obtain below 
if additional polarization information were included. This possibility will 
be considered elsewhere{\cite {inpro}}. 

The subprocess cross section for the unpolarized $\gamma \gamma \to f\bar f$ 
reaction including the contribution 
from graviton exchange can be written{\cite {pheno}} in a rather simple form:
\begin{equation}
{d\hat \sigma \over {dz}}={2\pi \alpha^2\over {\hat s}}N_c {1+z^2\over {1-z^2}}
\biggl[Q_f^2-\lambda K {\hat s^2(1-z^2)\over {4\pi \alpha M_s^4}}\biggr]^2\,,
\end{equation}
where as before $z=\cos \theta$ and $N_c$ is the usual color factor for the 
(assumed to be massless) fermions $f$. To obtain the true cross section 
integrated over a given angular bin, 
assuming that the two photons have a head-on collision, we must fold in the 
photon fluxes and integrate over them:
\begin{equation}
\sigma=\int^{x_{max}}~dx_1~\int^{x_{max}}~dx_2~\int_{bin}~dz~f_\gamma(x_1)
f_\gamma(x_2){d\hat \sigma \over {dz}}\,,
\end{equation}
where we explicitly identify $\hat s=s_{e^+e^-}x_1x_2$. The {\it lower} range 
of the above integrations requires some discussion. In principle, the photon 
fluxes persist to very low values of $x$; however, for very small $x$'s we lose 
significant sensitivity to $M_s$. Hence we want to maximize as 
much possible the luminosity of the flux with the greatest possible value of 
$\hat s^2/M_s^4$ as is easily seen by an examination of the equation above. 
To this end we impose the constraint that $\hat s/s \geq 0.01$ and also demand 
$x_{1,2}\geq 0.01$ subject to this constraint. As before we will impose a 
$10^o$ angular cut in our analysis in order to 
obtain our search reach as a function 
of the total $\gamma \gamma$ integrated luminosity. 
Additional cuts which, for example, balance the energy of the two incoming 
photons, are also possible but we do not make use of them here. 

In the case of $\gamma \gamma \to t\bar t$ production, the subprocess cross 
section is somewhat more cumbersome:
\begin{eqnarray}
{d\hat \sigma\over {dz}}&=&{d\sigma^{SM}\over {dz}}-{3\beta \over \hat s}
\left[{(\lambda K)^2\over\pi M_s^8}-Q_t^2 {2\alpha \lambda K
\over M_s^4(m_t^2-\hat t)(m_t^2-\hat u)}\right]\nonumber \\
& &\left[6m_t^8-4m_t^6(\hat t+\hat u)+4m_t^2 \hat t \hat u
(\hat t+\hat u)-\hat t\hat u(\hat t^2+\hat u^2)+m_t^4(\hat t^2+
\hat u^2-6\hat t \hat u)\right]\,,
\end{eqnarray}
with $\hat t,\hat u={-1\over 2}\hat s(1\mp \beta z)+m_t^2$, with 
$\beta^2 =1-4m_t^2/\hat s)$, which apart from color factors, agrees with the 
results of Mathews, Raychaudhuri and Sridhar{\cite {pheno}} for the cross 
section for $gg\to t\bar t$. 
In the present case the kinematics require the photon energies to satisfy the 
constraint 
$x_1x_2\geq 4m_t^2/s$ which then determines the lower bounds on $x_{1,2}$.

To get a rough idea of the sensitivity of $\gamma \gamma$ collisions to $M_s$ 
we display in Fig.8 the angular distribution for the case where $f$ is 
summed over light quarks (\ie, a final state of two jets 
without flavor tags) at a
500 GeV collider with a {\it diphoton} integrated luminosity of 100$~fb^{-1}$. 
Due to the SM $u$- and $t$-channel exchanges there is an enormous flux 
in both the forward and backward directions. However the true region of 
sensitivity is at large angles where the rate is the smallest as was the case 
for Moller scattering. Note that the deviations are easily distinguished from 
both the SM and each other. It is clear from this figure that the search 
reach for $M_s$ would again exceed 1.5 TeV independent of the choice of the 
sign of $\lambda$ if $\gamma \gamma \to f\bar f$ were the only relevant 
process. However, since in this case the final state fermions are not tagged,  
the process $\gamma \gamma \to gg$, which occurs only through the exchange 
of a K-K tower, would now also contribute since the 
final state in both cases is just two jets as far as a detector is concerned. 

\vspace*{-0.5cm}
\nn
\begin{figure}[htbp]
\centerline{
\psfig{figure=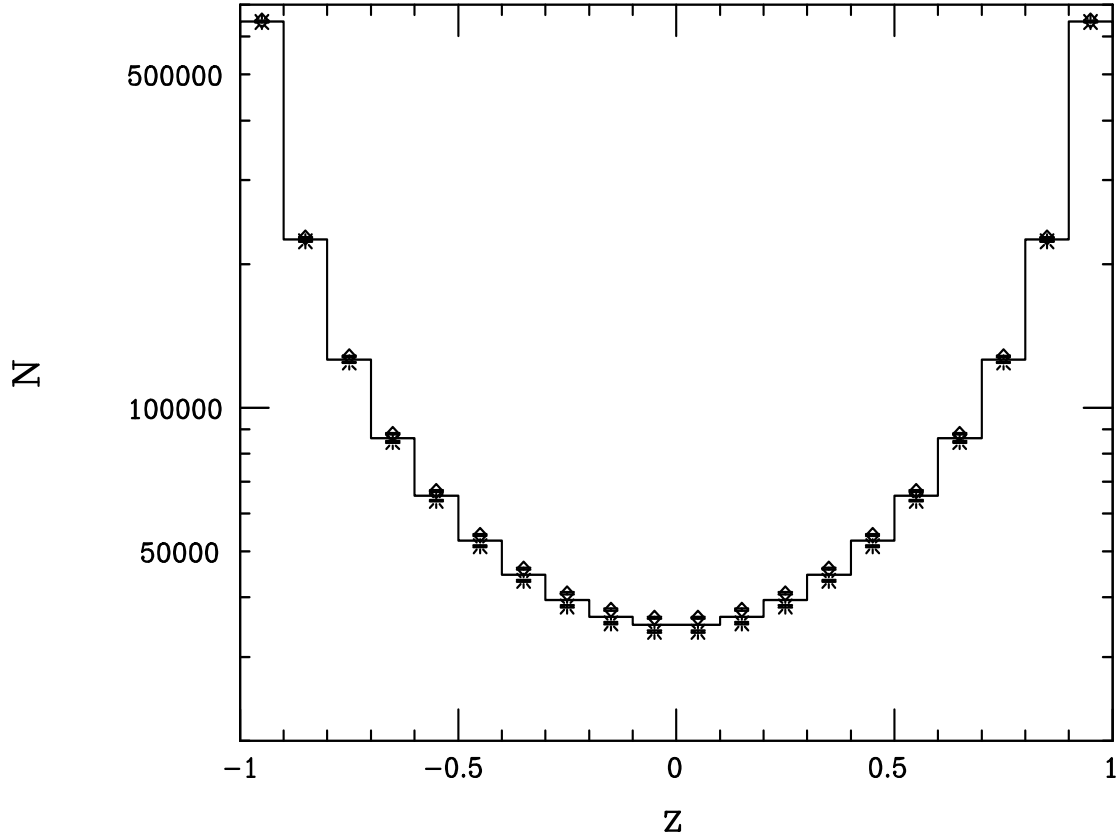,height=14cm,width=17cm,angle=-90}}
\vspace*{-1cm}
\caption[*]{Angular distribution for the process $\gamma \gamma \to q\bar q$, 
with $q$ being summed over the five light flavors of 
quarks at a 500 GeV $e^+e^-$ collider 
with an {\it integrated} photon luminosity of 100$~fb^{-1}$ assuming the cuts 
described in the text. The SM corresponds to 
the histogram while the `data' represent the 
ADD scenario with $M_s=$1.5 TeV for $\lambda=\pm$~1.}
\label{fig7}
\end{figure}
\vspace*{0.4mm}

To obtain the search reach there are thus two possibilities: first, one may 
add flavor tagging for the quarks $c$ and $b$ which removes the contribution 
from the $gg$ final state. Second, we may drop tagging and include the 
$gg$ contribution. Following the first approach and 
combining the $f=c,b,t$ final states together with $f=e$, $\mu$ and $\tau$, we 
proceed as above using the efficiencies of Hewett{\cite {pheno}}. In the 
second analysis, we add the contribution from $\gamma \gamma \to gg$ 
to that from all light quarks together with the leptons 
and follow a similar procedure. We note here that the $\gamma \gamma \to gg$ 
subprocess cross section due to graviton tower exchange takes the following 
simple form:
\begin{equation}
{d\hat \sigma \over {dz}}={\lambda^2K^2\hat s^3\over {32\pi M_s^8}}[1+6z^2
+z^4]\,,
\end{equation}
The results of these two different analyses are shown together in 
Fig.9 for $\sqrt {s_{e^+e^-}}$=500 GeV, 1 TeV and 1.5 TeV colliders. 
Here we see that for reasonable luminosities the search reach is 
$\simeq 4-5\sqrt {s_{e^+e^-}}$, 
which is impressive considering the minimum energy degradation of $\geq 17\%$ 
in going to the $\gamma \gamma$ center of mass frame. Relative to 
$\sqrt {s^{max}_{\gamma \gamma}}$ the search lies in the range of 
$\simeq 5-6\sqrt {s_{\gamma \gamma}}$ comparable to that 
found for either $e^+e^-$ or 
$e^-e^-$ collisions. As one would expect the reach obtained from the 
non-tagged analysis, which has greater statistics, is somewhat better but not 
by a very large amount. We note that the search reach does not increase as 
rapidly with $\sqrt s$ as does Bhabha and Moller scattering due to effects 
of the photon spectra. 
Again it is quite clear that the $M_s$ reach obtained 
from $\gamma \gamma$ collisions will greatly 
complement those resulting from $e^+e^-$ and $e^-e^-$ interactions.

\vspace*{-0.5cm}
\nn
\begin{figure}[htbp]
\centerline{
\psfig{figure=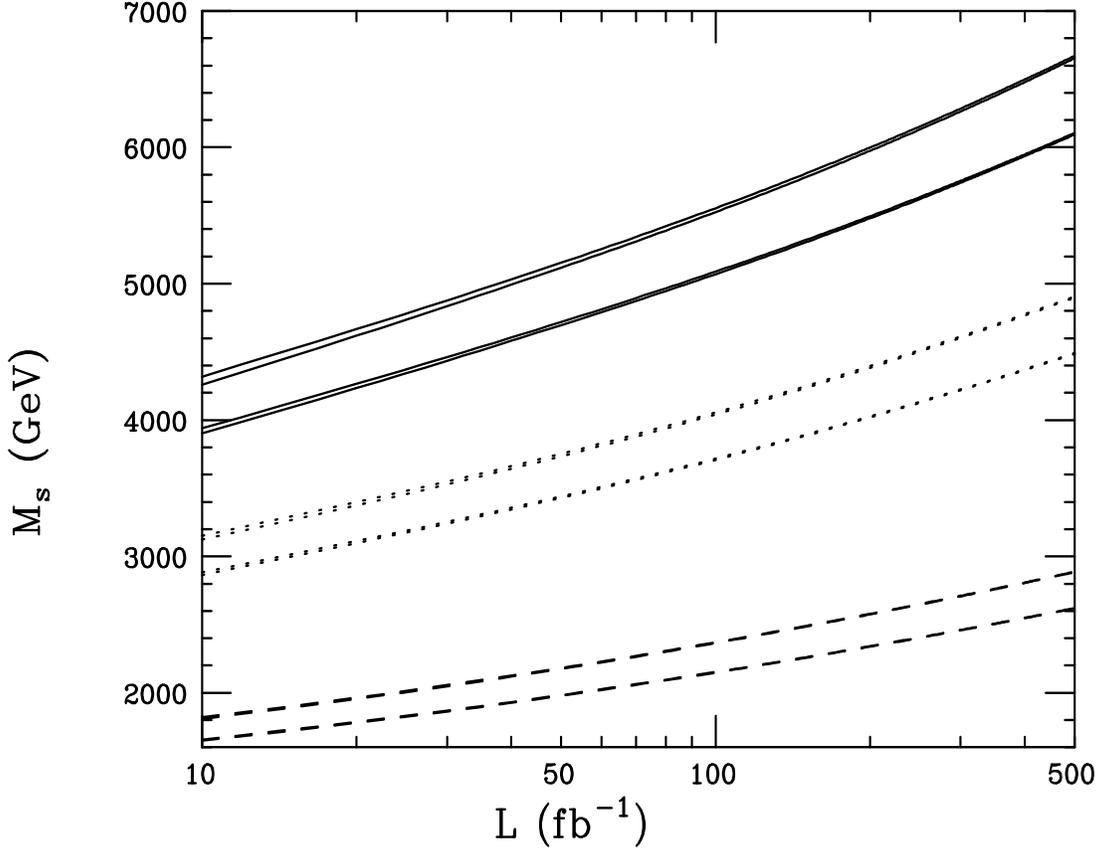,height=14cm,width=17cm,angle=-90}}
\vspace*{-1cm}
\caption[*]{Search reaches for the processes $\gamma \gamma \to f\bar f$, 
with $f$ being the $c,t$ and $b$ quarks together with $e$, $\mu$ and 
$\tau$(lowest curve of a given type), and for lepton pairs, top, plus 
light quark jets(upper pair of 
curves) as a function of the total $\gamma \gamma$ integrated luminosity. At a 
500(1000, 1500) GeV $e^+e^-$ collider the result is given by 
the dashed(dotted, solid) curve and in the former case is 
essentially independent of the choice $\lambda=\pm$ 1. The details of the 
analysis are described in the text.}
\label{fig8}
\end{figure}
\vspace*{0.4mm}

\section{Summary and Conclusions}

In this paper we have extended the phenomenological analyses of the ADD 
scenario presented in Ref.5 to a number of new processes involving the 
exchange of a Kaluza-Klein tower of gravitons at various types of colliders. 
The main points of our analysis are as follows: 
\begin{itemize}

\item  The collection of approximately 1 $fb^{-1}$ of integrated luminosity at 
HERA balanced equally between the four intial charge and polarization 
states, $e^\pm_{L,R}$, will lead to a $95\%$ CL bound on the values of $M_s$ 
in excess of 1 TeV. This bound is comparable to that obtainable at Run II of 
the Tevatron employing the Drell-Yan process and that derivable by combining 
the results of the four LEP experiments after all data taking is completed. 
Clearly, the measurements at all three colliders are complementary. We 
estimate the current lower bound on $M_s$ from existing HERA data using an 
unpolarized $e^+$ beam at a lower center of mass energy to be no more than 
$\simeq 500-600$ GeV.

\item  Low energy $\nu N$ scattering data, while of high precision, are not 
able to significantly constraint the value of $M_s$ although the same data 
is known to place respectable constraints on dimension-six operators 
associated with conventional contact interactions arising due to 
compositeness. This lack of sensitivity is directly related to the fact that 
the K-K tower exchange leads to dimension-eight operators which are thus 
suppressed by more than three orders of magnitude in comparison to contact 
interactions. The high precision of these measurements do not compensate in 
this case for the low energy at which they are made. 

\item  Both Bhabha and Moller scattering were shown to have comparable 
sensitivity to the exchange of K-K towers of gravitons with search reaches of 
the same magnitude as those obtained by Hewett{\cite {pheno}} for the more 
conventional $e^+e^-\to 
f\bar f$ process, \ie, $\simeq 6\sqrt s$. The behavior of the search reach 
for these two processes with variations of integrated luminosity were, 
however, quite different due to the relative importance of systematic errors. 
This is due to the large cross sections for Bhabha and Moller scattering 
resulting from QED poles in the forward(and backward for Moller scattering) 
directions even after acceptance cuts are applied. 

\item  The $\gamma \gamma \to f\bar f$ is a particularly clean channel for 
new physics without electroweak contributions at tree level beyond QED. 
In addition, there are no gauge invariant dimension-six operators arising 
from contact interactions in this case. As in the case of both Bhabha and 
Moller scattering cross sections are very large due to both $t$- and $u$-
channel poles and systematic effects are important in setting limits. In 
comparison to $e^-e^\pm$ reactions, $\gamma \gamma$ reactions suffer in their 
$M_s$ reach due to the reduced effective center of mass energy induced by the 
continuous photon spectrum from the backscattered laser. However we found that 
by summing over all leptons as well as all light quark flavors and gluon pairs 
the search reach for $M_s$ could be as 
large as $5\sqrt s$ which is quite comparable to the $e^-e^\pm$ searches 
and quite complementary. The use of photon beam polarization may 
lead to an increase in this search reach{\cite {inpro}}. 

\item  Signals for an exchange of a Kaluza-Klein tower of gravitons in the 
ADD scenario of low energy quantum gravity appear in many complementary 
channels 
{\it simultaneously} at various colliders. Such signatures for new physics  
are rather unique and will not be easily missed. 

\end{itemize}
The discovery of new dimensions may be at our doorstep and may soon make 
their presence known at existing and/or future colliders. Such a discovery 
would revolutionize the way we think of physics beyond the electroweak scale.

\noindent{\Large\bf Acknowledgements}

The author would like to thank J.L. Hewett, N. Arkani-Hamed, J. Wells, T. Han 
and J. Lykken for discussions related to this work.

Note added: After the present analysis was completed we received a paper by 
Mathews, Raychaudhuri and Sridhar{\cite {mrs}} who considered the present 
bounds on the scale $M_s$ from HERA data. Their resulting bound is in 
qualitative numerical agreement with that obtained in the discussion above.

\newpage

%
\def\MPL #1 #2 #3 {Mod. Phys. Lett. {\bf#1},\ #2 (#3)}
\def\NPB #1 #2 #3 {Nucl. Phys. {\bf#1},\ #2 (#3)}
\def\PLB #1 #2 #3 {Phys. Lett. {\bf#1},\ #2 (#3)}
\def\PR #1 #2 #3 {Phys. Rep. {\bf#1},\ #2 (#3)}
\def\PRD #1 #2 #3 {Phys. Rev. {\bf#1},\ #2 (#3)}
\def\PRL #1 #2 #3 {Phys. Rev. Lett. {\bf#1},\ #2 (#3)}
\def\RMP #1 #2 #3 {Rev. Mod. Phys. {\bf#1},\ #2 (#3)}
\def\ZPC #1 #2 #3 {Z. Phys. {\bf#1},\ #2 (#3)}
\def\EJPC #1 #2 #3 {E. Phys. J. {\bf#1},\ #2 (#3)}
\def\IJMP #1 #2 #3 {Int. J. Mod. Phys. {\bf#1},\ #2 (#3)}

\end{document}